\documentclass[preprint,aps]{revtex4}
\usepackage{graphicx}
\usepackage{psfrag}
\usepackage{color}

\begin{document}
\bibliographystyle{prsty}

\title{Plastic response of a 2d amorphous solid to quasi-static shear :\\ 
I - Transverse particle diffusion and phenomenology of dissipative events}
\author{Ana\"el Lema\^{\i}tre$^{(1)}$}
\author{Christiane Caroli$^{(2)}$}
\affiliation{$^{(1)}$ Institut Navier-- LMSGC, 2 all\'ee K\'epler,77420 Champs-sur-Marne, France}
\affiliation{$^{(2)}$ INSP, Universit\'e Pierre et Marie Curie-Paris 6, 
CNRS, UMR 7588, 140 rue de Lourmel, 75015 Paris, France}

\date{\today}

\begin{abstract}
We perform extensive simulations of a 2$D$ LJ glass subjected to quasi-static shear 
deformation at $T=0$.
We analyze the distribution of non-affine displacements in terms of contributions
of plastic, irreversible events, and elastic, reversible motions.
From this, we extract information about correlations between plastic events and about
the elastic non-affine noise. Moreover, we find that non-affine motion is essentially diffusive,
with a clearly size-dependent diffusion constant. These results, supplemented
by close inspection of the evolving patterns of the non-affine tangent displacement field,
lead us to propose a phenomenology of plasticity in such amorphous media.
It can be schematized in terms of elastic loading and irreversible flips of small, 
randomly located shear transformation zones, elastically 
coupled via their quadrupolar fields.
\end{abstract}

\maketitle

\section{Introduction}
Plastic deformation of amorphous solids and, more generally, of jammed 
disordered media (foams, confined granular media, colloidal 
glasses,\ldots) has been intensively studied in the past two decades. 
General agreement is now gradually emerging about the 
nature of the elementary dissipative events in these highly 
multistable systems. They consist in sudden rearrangements of 
small clusters comprising a few basic structural units, such as 
$T_{1}$ events in dry foams. In the case of glasses, where they cannot 
be observed directly, progress has come, following the pioneer work of 
Argon and collaborators, from extensive numerical studies~\cite{KobayashiMaedaTakeuchi1980,DengArgonYip1989,ArgonBulatovMottSuter1995}.
More recently, simulations performed on model systems -- 
Lennard-Jones (LJ) glasses -- have proved very helpful to improve 
our understanding of the effect of topological disorder on the elastic 
as well as plastic shear response of amorphous solids.

\begin{figure}[p]
\begin{center}
\psfrag{sigma}{{\large $\sigma$}}
\psfrag{gamma}{{\large$\gamma$}}
\includegraphics[width=0.45\textwidth,clip]{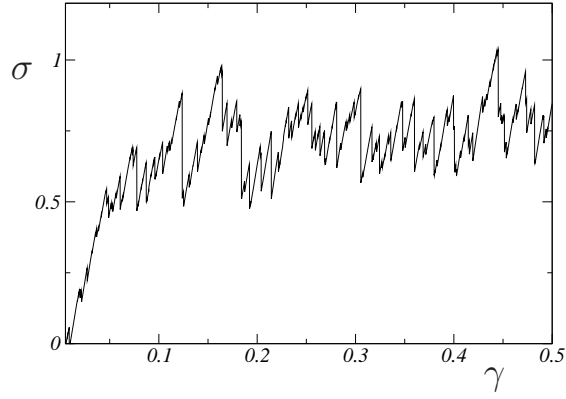}
\caption{\label{fig:stressstrain}
Stress-strain curve for a $20\times20$ system.
}
\end{center}
\end{figure}

MD simulations are instrumental in elucidating the thermal dependence 
of the flow stress $\sigma (\dot{\gamma})$ in the high strain rate 
($\dot{\gamma} >> 1$) regime. However, such conditions (finite $T$, 
high $\dot{\gamma}$) "blur" the microscopic motion, making it 
difficult to characterize precisely the elementary events, which are 
the building blocks on which constitutive laws should be based.
For this purpose, a second class of numerical works have focussed on 
the athermal ($T = 0$), quasi-static ($\dot{\gamma} \rightarrow 0$) 
regime. In this later regime, hereafter abbreviated as AQS, when a sample is sheared at constant 
rate, the stress-strain $\sigma 
(\gamma)$ curve exhibits (see Figure~\ref{fig:stressstrain}) elastic branches interrupted 
by discontinuous drops $\Delta\sigma$ which are the signature of the 
dissipative events. Beyond an initial transient, 
$\sigma(\gamma)$ fluctuates about an average value $\bar{\sigma}$, 
which is identified with the yield stress $\sigma_{Y} = 
\lim\limits_{\dot{\gamma}\to 0} \bar{\sigma}(\dot{\gamma})$. The 
distribution of stress drops is broad and system-size dependent.
In their study on a 2D LJ glass, Maloney and Lemaitre~\cite{MaloneyLemaitre2006} were 
able to analyze them in terms of cascades of elementary events, which 
we will term "flips". Each such flip involves both the strong 
rearrangement of a small cluster ( $\sim$ a few atoms), and the 
appearance of an associated quadrupolar elastic field. This result 
substantiates  the representation of elementary processes as 
Eshelby-like~\cite{Eshelby1957} shear transformations~\cite{PicardAjdariLequeuxBocquet2004}.

However, it appears desirable to analyze AQS simulations in more detail, 
in order to shed light upon debated questions concerning phenomenologies 
based on the notion of shear transformation zones (STZ)~\cite{BulatovArgon1994a,ArgonBulatovMottSuter1995,FalkLanger1998}.
Namely:

-- Can the flipping clusters be associated with regions of the disordered solid which retain
their identity over a finite range of elastic loading before they reach their instability 
threshold?

-- If so, do the simulations give information about the response of a zone to the elastic field 
generated by the flip of another zone?

-- Can one evaluate the relative importance of the dynamic noise resulting from this 
mechanism as compared with disorder-induced fluctuations of the non-affine elastic field?

In order to address these issues, we extend in this article a recent study 
on a 2D LJ glass~\cite{TanguyLeonforteBarrat2006}, by Tanguy et al, which
confirms that plastic flow is spatially heterogeneous. They 
claim that one should distinguish between two types of plastic 
events: strongly localized ones occurring during the initial loading 
phase ($\sigma << \sigma_{Y}$), and non-localized ones which they 
term "non-permanent shear bands".
They also investigate atomic motion in the direction transverse to 
the plastic flow. While they find it to be diffusive at long times, 
they conclude to its hyperdiffusive nature on short time intervals.

Transverse displacements are purely non-affine and, as indicated by 
the jagged shape of the $\sigma(\gamma)$ curve, consist of a succession of 
possibly noisy elastic episodes interspersed with sudden  
atomic rearrangements associated with the plastic events, hence the interest of a 
detailed analysis of their dynamics.
With this remark in mind, we revisit in Section II the analysis of 
transverse particle motion, now explicitly separating elastic and 
plastic contributions. We thus extract information about {\it (i)} 
correlations between plastic events {\it (ii)} the elastic 
contribution to the non-affine noise invoked in the phenomenological 
STZ model of Falk and Langer~\cite{FalkLanger1998}. Moreover, we find it 
inappropriate to qualify global  
transverse particle motion as hyperdiffusive at short times.
Indeed, the effective diffusion coefficient 
$ D (\Delta\gamma) = \langle\Delta y^{2}\rangle/\Delta 
\gamma$ smoothly increases with $\Delta \gamma$ from a {\it finite} 
short time value $D_{0}$ to the asymptotic value $D_{\infty}$.
Noticeably, the whole $D(\Delta \gamma)$ curve is {\it system size 
dependent} -- a result which appears consistent with the analysis of 
plastic events in terms of flip cascades. 
We also show that the contribution of plastic events to transverse diffusion dominates
markedly over the effect of disorder-induced non-affine elastic fluctuations.

In Section III, on the basis of close inspection of the evolution 
of the spatial structure of the infinitesimal non-affine field, we 
show that our results can be interpreted in terms of the elastic loading of zones driven
by shear, which soften gradually as they approach their spinodal limit.
As they approach this threshold, these zones give rise to quadrupolar fields of growing 
amplitude. 
The zones thus identified can be traced back over shear intervals substantially larger
than the average interval between plastic events, the effect of which is thus easily observable.

These observations support a description in terms of elastically loaded zones,
and point towards the relevance of the dynamical noise generated by plastic flips 
themselves.
The associated inter-zone elastic couplings should be responsible for the "autocatalytic 
avalanches"~\cite{DemkowiczArgon2005,BaileySchiotzLemaitreJakobsen2007} 
or flip cascades~\cite{MaloneyLemaitre2004,MaloneyLemaitre2006}
constituting the system-size dependent plastic 
events, which we believe to be precisely the non-permanent shear bands 
invoked in ref.\cite{TanguyLeonforteBarrat2006}.
We attribute the stronger localization 
of initial events to the smaller density of nearly unstable zones in 
the as-quenched or weakly stressed samples.

This empirical study thus leads to the emergence of a phenomenology of 
the non-affine shear response summarized in Section IV.
While supporting the concept of shear transformation zones (STZ) or, equivalently, 
elastically loaded traps (SGR), it diverges from these models about 
two of their basic assumptions, namely 
independence of elementary events and the nature of noise.
We think that it should be of use for further 
developments in the modelization of plasticity of amorphous media.

\section{Transverse particle motion}

\subsection{2D simulations in the AQS regime}
We use here the same binary LJ mixture as that of ref.\cite{MaloneyLemaitre2006},
namely large (L) and small(S) particle radii and numbers are: 
$r_{L} = 0.5$, $r_{S} = 0.3$, and $N_{L} = N_{S} (1+ \sqrt{5})/4$.
These values ensure that no crystallization occurs at low temperature.
Simple shear deformation is imposed using Lees-Edwards boundary 
conditions. We study systems of three different sizes $L \times L$, 
with $L = 10, 20, 40$.

The quasi-static regime corresponds to the limit where the external 
time scale $\dot{\gamma}^{-1}$ is much larger than that of internal 
relaxation processes.
The system, starting from local equilibrium, thus 
follows adiabatically the shear-induced evolution of the 
corresponding energy minimum up to the spinodal limit where this 
minimum disappears and the system jumps into another local minimum of 
lower energy. The simulation proceeds as follows: a small increment 
$\delta$ of homogeneous shear strain is imposed, then energy is 
minimized using a conjugate gradient algorithm with a stringent 
convergence criterion (see~\cite{MaloneyLemaitre2006} for details). 
We choose $\delta = 10^{-4}$, small 
enough to ensure that, for our system sizes, all elastic branches 
are well resolved. A typical $\sigma(\gamma)$ curve is displayed on 
Figure~\ref{fig:stressstrain}.

Starting from an initial quench, we explore the shear range $\gamma 
\leq 2$. In order to characterize the stationary state, we only retain 
data for $\gamma > 0.1$, which ensures that initial transients are 
discarded, thus making details of the quenching protocol immaterial.
We have been able to collect data on $100$, $50$ and $20$ 
systems of respective sizes $ L = 10, 20, 40$.

Following ref.\cite{MaloneyLemaitre2006}, we make extensive use of the so-called 
non-affine tangent field $\lbrace \bf{u}_{i}\rbrace$, 
defined as the difference between the linear response of particle 
displacements to an increment of homogeneous strain and the 
corresponding homogeneous field. It is 
well-defined, and is computed, everywhere along each elastic branch.

\subsection{Statistics of transverse displacements}
In plastically deforming amorphous systems, non-affine 
displacement fields contain information about both, 
departures from standard continuum elastic behavior
and the nature of plastic events.
In our simple shear geometry, transverse particle 
displacements $\Delta y_{i}$ are purely non-affine, while longitudinal 
ones mix affine and non-affine contributions. We thus focus on the 
normalized distributions $P(\Delta y, \Delta\gamma)$ of the $\Delta 
y_{i}$'s for a fixed strain interval $\Delta\gamma$. The statistical 
ensemble is built by sampling, for all initial configurations, all 
the $\Delta y_{i} = y_{i}(\gamma_{0} + \Delta\gamma) - 
y_{i}(\gamma_{0})$ at each step $\gamma_{0}$ ($0.1 \leq \gamma_{0} \leq 
2 - \Delta\gamma$).
\begin{figure}[p]
\begin{center}
\psfrag{p}{{$P$}}
\psfrag{y}{{$\Delta y$}}
\psfrag{pzoom}{{$P$}}
\psfrag{yzoom}{{$\Delta y$}}
\includegraphics[width=0.45\textwidth,clip]{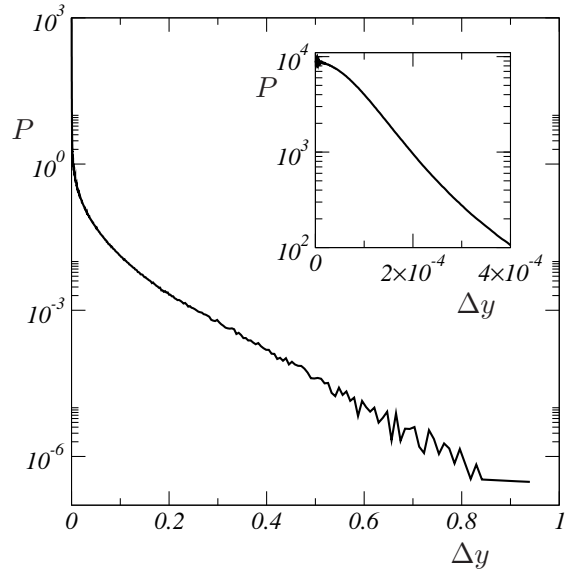}
\caption{\label{fig:2}
The distribution $P(\Delta y, \delta)$ of transverse particle displacements for 
the elementary strain step $\delta$ and for system size $L=20$.
Insert: Blow-up of the small $\Delta y$ region. 
}
\end{center}
\end{figure}

\begin{figure}[p]
\begin{center}
\psfrag{p}{{$\tilde P$}}
\psfrag{z}{{$\zeta$}}
\includegraphics[width=0.45\textwidth,clip]{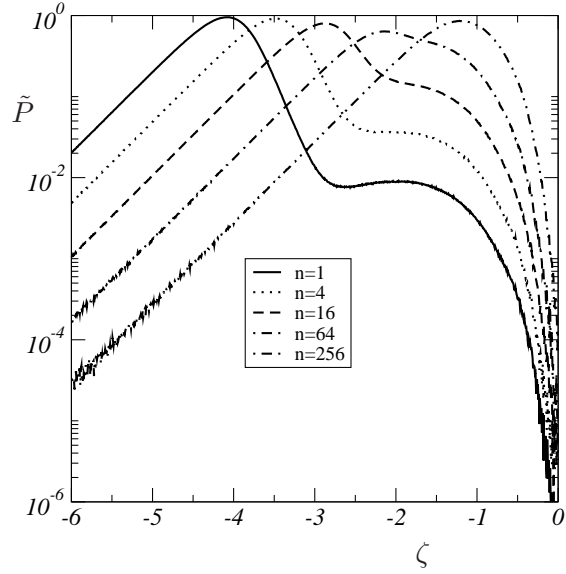}
\caption{\label{fig:3}
The distribution $\tilde{P}(\zeta; \Delta\gamma)$
of the scale variable $\zeta = \log (\Delta y)$ for increasing values of
$n=\Delta\gamma / \delta =1,4,16,64,256$. System size: $L=20$.
}
\end{center}
\end{figure}

Our results agree qualitatively with those of Tanguy et al~\cite{TanguyLeonforteBarrat2006}. Namely (see Figure~\ref{fig:2}) $P(\Delta y, \delta)$ has a 
quasi-gaussian center and exhibits, 
beyond $\Delta y \sim 0.2$, an exponential tail. Its fine structure 
shows more clearly in the log-lin representation of the distribution 
$\tilde{P}(\zeta)$ of the scale variable $\zeta = {\rm log}(\Delta y)$. 
As displayed on Figure~\ref{fig:3}, for small strain intervals 
($\Delta\gamma / \delta \sim$ a few units), the corresponding curves 
exhibit a peak at small $\Delta y$ in the $10^{-4}$ range, together 
with a broad hump for larger values. As $\Delta\gamma$ 
increases, the peak shifts to the right and the hump amplitude grows, 
though without noticeable
 horizontal shift, until both merge, for $\Delta\gamma \approx 
10^{-2}$.

\begin{figure}[p]
\begin{center}
\psfrag{p}{{$\tilde P$}}
\psfrag{z}{{$\zeta$}}
\includegraphics[width=0.45\textwidth,clip]{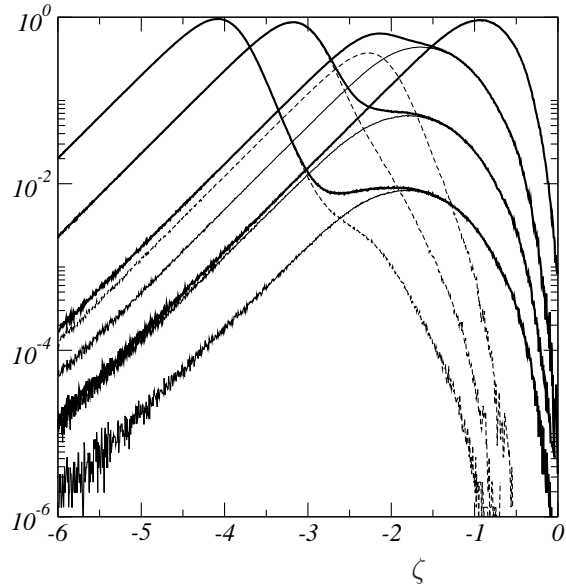}
\caption{\label{fig:4}
Decomposition of the total distributions $\tilde{P}(\zeta; \Delta\gamma)$ (thick solid lines),
for $n=\Delta\gamma / \delta =1,8,64,512$. For increasing $n$'s, 
the maximum of the distribution shifts rightwards.
Thin solid lines: contribution of plastic events;
thin dashed lines: contribution of elastic branches (see text).
System size $L=20$.
}
\end{center}
\end{figure}

Can we interpret this structure and its evolution in the light of the 
succession of elastic and plastic episodes which reflect into the 
sawtooth shape of the $\sigma (\gamma)$ response? Clearly, particle 
motion consists of a series of continuous trajectories interrupted by 
sudden jumps associated, respectively, with elastic episodes and 
plastic events. Since $\delta$ 
is our numerical strain resolution, each interval of length $\delta$ 
is, within our accuracy, defined as either purely elastic or 
purely plastic. So, for $\Delta\gamma = \delta$, we can decompose $P$ 
(and likewise $\tilde P$) as :
\begin{equation}
\label{eq:decomp}
P(\Delta y, \delta) = \alpha_{pl}(\delta) \Pi_{pl}(\Delta y, 
\delta) + (1 - \alpha_{pl}(\delta)) \Pi_{el}(\Delta y, \delta)
\end{equation}
where $\alpha_{pl} (\delta)$ is the fraction of "plastic intervals" 
in our ensemble, and $\Pi_{pl}$ (resp.  $\Pi_{el}$) are the normalized 
distributions associated with the plastic (resp. elastic) sub-ensemble. 
It now clearly appears (see Figure~\ref{fig:4}) that the small $\Delta y$ 
behavior of $P(\Delta y, \delta)$ results entirely from elastic 
motion, while its hump and large-$\Delta y$ tail are due to 
plastic events.

On this basis, we are now able to understand the evolution of $\tilde{P}$ 
with $\Delta\gamma$. We extend the above decomposition to 
$\Delta\gamma > \delta$, now defining as "plastic" any interval 
containing at least one plastic event. As $\Delta\gamma$ increases 
while remaining small with respect to the average length 
($\delta \alpha_{pl}(\delta)^{-1})$) of an elastic branch, most plastic 
intervals in general still contain only one plastic event, so that 
their fraction $\alpha_{pl}(\Delta\gamma) \approx (\Delta\gamma / 
\delta) \alpha_{pl}(\delta)$, while $\Pi_{pl}$ remains quasi 
unchanged, because the scale of plastic slips is large compared with 
that of elastic ones. This explains the evolution (Figure~\ref{fig:4}) of the plastic hump 
which, as $\Delta\gamma$ increases, consists primarily 
in an upward shift with little change in shape. 

\begin{figure}[p]
\begin{center}
\psfrag{p}{{$\tilde{\Pi}_{el}$}}
\psfrag{y}{{$\Delta y/\Delta\gamma$}}
\includegraphics[width=0.45\textwidth,clip]{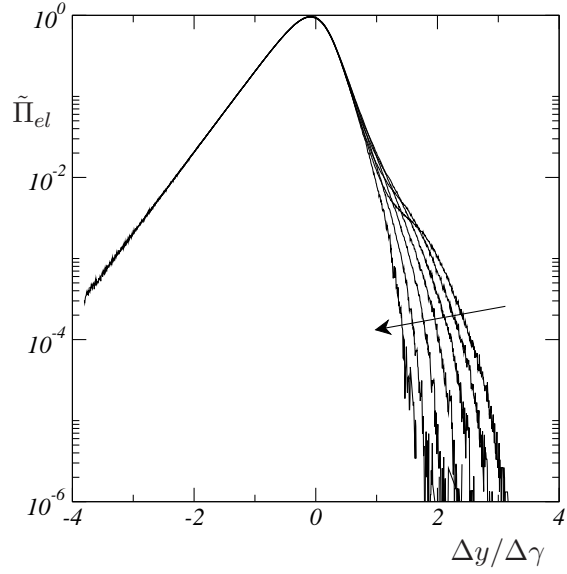}
\caption{\label{fig:5}
The function $\tilde{\Pi}_{el}$ plotted versus $\Delta y/\Delta\gamma$
for $n=\Delta\gamma / \delta =1,2,4,8,16,32,64$.
The arrow indicates the direction of {\it increasing} $n$'s.
System size $L=20$.
}
\end{center}
\end{figure}

In this regime, which  
corresponds to $\Delta\gamma / \delta << \alpha_{pl}(\delta)^{-1}$ ( 
$\sim 100$ for the $20 \times 20$ system), the fraction 
$\alpha_{el} = (1 - \alpha_{pl})$ of purely elastic intervals remains 
nearly constant (e.g., for $L = 20$ and $\Delta\gamma = 10 \delta$, 
$\delta\alpha_{el}/\alpha_{el} \sim 0.1$), so that the variations 
with $\Delta\gamma$ of the elastic part of $\tilde{P}$, 
$\tilde{P}_{el} = \alpha_{el}(\Delta\gamma) \tilde{\Pi}_{el}$, 
directly reveal those of $\tilde{\Pi}_{el}$. Inspection of the 
numerical data suggests (see also Figure~\ref{fig:4}) that the rightward shifts 
of $\tilde{P}_{el}(\zeta)$ are roughly equal to ${\rm log}(\Delta\gamma)$. 
The plot (Figure~\ref{fig:5}) of $\tilde{P}_{el}$ {\it vs} 
$\log(\Delta y/\Delta\gamma$)) shows that the collapse is indeed 
excellent in the low and middle ranges of $(\Delta y/\Delta\gamma)$ 
values, but fails in the large $\Delta y$ tails.

This collapse means that {\it during the purely elastic episodes separating 
plastic events non-affine displacements are essentially convective}:
the $\gamma$-derivative of the $y_i$ field is nearly constant
over most of an elastic episode.
This amounts to stating that the tangent non-affine 
field $\lbrace\bf{u}_{i}\rbrace$, although spatially disordered~\cite{footnote}, 
remains quasi-quenched.
Now, we know from ref.~\cite{MaloneyLemaitre2004a} that this cannot be true when the 
system comes close below a spinodal threshold $\gamma_{c}$. Indeed, in these 
near-critical regions, a gradually softening elastic mode develops, leading to 
a $(\gamma_{c} - \gamma )^{-1/2}$ divergence of $\lbrace\bf{u}_{i}\rbrace$. 
Clearly, it is 
the contribution of these near-critical softened configurations which 
explains the large-$\Delta y$ tails of $\tilde{\Pi}_{el}$. The tail 
deflation (Figure~\ref{fig:5}) results from the fact that, as the interval 
$\Delta\gamma$ increases, the weight of these soft configurations is 
gradually transferred from $\tilde{\Pi}_{el}$ to $\tilde{\Pi}_{pl}$.

This analysis clarifies the physical significance of the shape of 
$P(\Delta y, \Delta\gamma)$ in the moderate $\Delta\gamma$ range 
(where its non-Gaussian parameter remains large~\cite{TanguyLeonforteBarrat2006}): 
{\it (i)} Its quasi-gaussian center~\cite{footnote} 
results from small-scale non-affine displacements accumulated along 
purely elastic segments, during which particle trajectories are 
essentially convected. {\it (ii)} Its quasi-exponential tail arises 
from plastic jumps during plastic events.

For larger $\Delta\gamma/\delta \gtrsim \alpha_{pl}(\delta)^{-1}$, however,
the decomposition of equation (\ref{eq:decomp}) loses physical content since
the statistical weight of elastic intervals vanishes while $\tilde{\Pi}_{pl}$
mixes both elastic and plastic contributions.
In order to circumvent this limitation, we concentrate in the following on the 
evolution with $\Delta\gamma$ of the second moment $\langle \Delta y^{2}\rangle$.

\subsection{Transverse particle diffusion}
\begin{figure}[p]
\begin{center}
\psfrag{D}{{$D$}}
\psfrag{deltagamma}{{$\Delta \gamma$}}
\includegraphics[width=0.45\textwidth,clip]{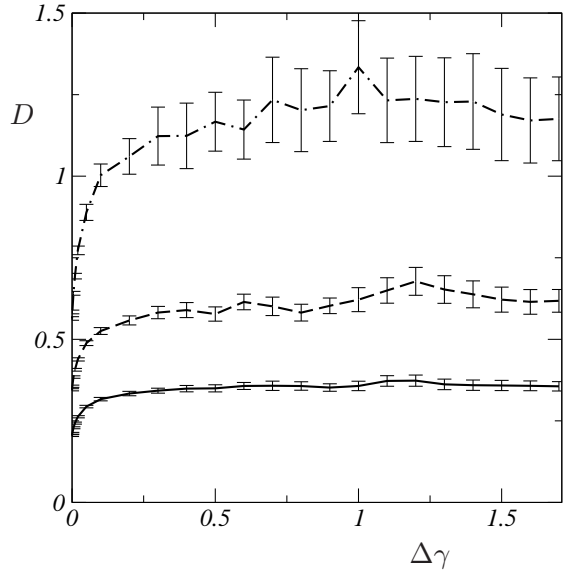}
\caption{\label{fig:6}
From bottom to top: the instantaneous diffusion constant, 
$D(\Delta\gamma) = \langle \Delta y^{2}\rangle/\Delta\gamma$ versus $\Delta\gamma$,
for increasing system sizes $L=10,20,40$ (and respectively 100, 50, and 20 samples).
Error bars result from a standard Student analysis of the data.
}
\end{center}
\end{figure}

In order to elucidate the nature of the transverse particule dynamics, 
we have computed from our data the space and 
ensemble average $\langle \Delta y^{2}\rangle$ for increasing values 
of $\Delta\gamma$. On Figure~\ref{fig:6} we plot the effective diffusion 
coefficient $D(\Delta\gamma) = \langle \Delta 
y^{2}\rangle/\Delta\gamma$. For our three system sizes, 
$D(\Delta\gamma)$ exhibits the same qualitative features, namely 
it increases from a finite value $D_{0} = \lim\limits_{\Delta\gamma\to 0} D$, 
and saturates at a finite value $D_{\infty}$ for large $\Delta\gamma$ 
($\gtrsim 0.5$). However, $D$ is strikingly system-size dependent, 
$D^{(L)}(\Delta\gamma)$ increasing with $L$ for all $\Delta\gamma$'s.

Again, in order to shed light on the origin of this behavior, let us 
separate explicitly the elastic and plastic contributions to non-affine 
particle motion. Indeed, we can write: 
\begin{equation}
\label{eq:traj}
\frac{dy_{i}}{d\gamma} = u_{iy}(\gamma) + \sum_{a} Y_{i}^{a} \delta 
(\gamma - \gamma_{a})
\end{equation}
where $a$ labels the strain values where irreversible events 
(avalanches) occur, so that :
\begin{equation}
\label{eq:corr}
D(\Delta\gamma) = D_{ee}(\Delta\gamma) +  
D_{ep}(\Delta\gamma) +  D_{pp}(\Delta\gamma)
\end{equation}
with
\begin{equation}
 \label{eq:ee}
 D_{ee}(\Delta\gamma) = \frac{1}{\Delta\gamma}\left\langle 
 \int_{\gamma_{0}}^{\gamma_{0} + 
 \Delta\gamma}\, d\gamma \int_{\gamma_{0}}^{\gamma_{0} + 
\Delta\gamma} \,d\gamma' u_{iy}(\gamma) u_{iy}(\gamma ') \right\rangle
\end{equation}
and $D_{ep}$, $D_{pp}$ the corresponding cross-correlated and 
plastic-plastic contributions. These three functions carry the 
information about self and cross correlations of elastic and plastic 
non-affine displacements.

Let us first consider the $\Delta\gamma\to 0$ limit. A straightforward 
asymptotic analysis taking into account the square-root divergence of
the tangent field $\lbrace\bf{u}_{i}(\gamma)\rbrace$ at spinodal points 
leads to:
\begin{equation}
\label{eq:limep}
D_{ep}(\Delta\gamma) = {\cal O}(\Delta\gamma^{1/2})
\end{equation}

\begin{equation}
\label{eq:limee}
D_{ee}(\Delta\gamma) = {\cal O}\left( - 
\Delta\gamma\,{\rm Log}(\Delta\gamma)\right)
\end{equation}
So, their contribution to $D(\Delta\gamma)$ vanishes, and the finite 
value of $D_{0}$ originates only from avalanches:
\begin{equation}
\label{eq:D0}
D_{0} = \lim\limits_{\Delta\gamma\to 0} D_{pp}\,=\,f\,\left\langle 
\left(Y_{i}^{a}\right)^{2}\right\rangle
\end{equation}
with
\begin{equation}
\label{eq:f}
f = \lim\limits_{\delta\to 
0}\left[\frac{\alpha_{pl}(\delta)}{\delta}\right]
\end{equation}
the average avalanche frequency, and 
$\langle \left(Y_{i}^{a})^{2}\right)\rangle$ the variance of 
transverse displacements in a single plastic event.

It was shown in~\cite{MaloneyLemaitre2004,MaloneyLemaitre2006} 
that $f$ increases with size roughly as the 
lateral system size $L$. Using the measured frequency, we find that 
the variance $\langle \left(Y_{i}^{a})^{2}\right)\rangle = D_{0}f^{-1}$ 
decreases slowly with size: for $L = 10, 20, 40$, we get $10^{4}
\langle \left(Y_{i}^{a})^{2}\right)\rangle = 37, 26, 20$. We will 
come back to this point later.

\begin{figure}[p]
\begin{center}
\psfrag{D}{{$D$}}
\psfrag{deltagamma}{{$\Delta\gamma$}}
\includegraphics[width=0.45\textwidth,clip]{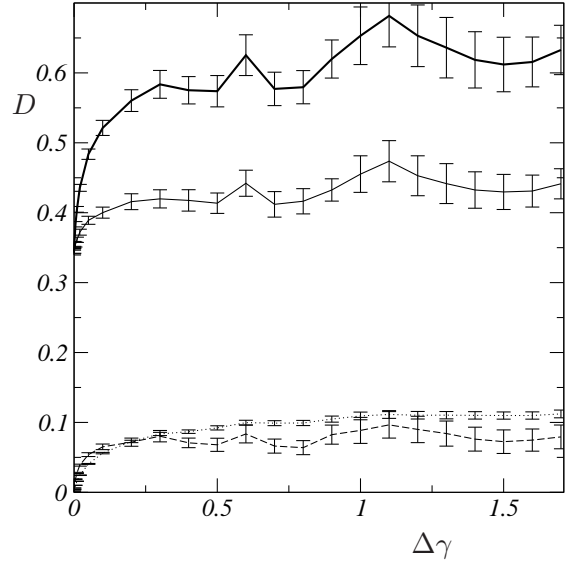}
\caption{\label{fig:7}
The instantaneous diffusion constant,
$D(\Delta\gamma) = \langle \Delta y^{2}\rangle/\Delta\gamma$
for system size $L=20$ (thick solid line) and its decomposition into
$D_{ee}$ (dotted line), $D_{ep}$ (dashed line) and $D_{pp}$ (thin solid line).
}
\end{center}
\end{figure}

Figure~\ref{fig:7} shows the decomposition (eq.(\ref{eq:corr})) for $L = 20$. It 
is seen that, for all $\Delta\gamma$, $D_{pp}$ remains the dominant 
contribution. $D_{pp}(\Delta\gamma)$ grows, then saturates. Its 
growth range $\Gamma_{pp} \sim 0.25$ shows no clear size dependence. 
This behavior entails that plastic events are correlated over a finite 
$\Delta\gamma$ range, which we understand as measuring a typical shear range over 
which the structural features involved in plastic events retain their 
identity.

For all sizes, $D_{ep}(\Delta\gamma)$, though much smaller than 
$D_{pp}$, exhibits an analogous behavior with a similar correlation 
range. This indicates that the above-mentioned persistent structures 
dominate the non-affine elastic response. Moreover the near 
square-root behavior~\cite{footnote2} of $D_{ep}$ for small $\Delta\gamma$ signals that 
the field $\lbrace\bf{u}_{i}\rbrace$ in a given near-critical region 
is strongly correlated with the displacements associated with the 
subsequent avalanche.

Finally, $D_{ee}$ also rises and saturates, but more slowly than $D_{ep}$ and $D_{pp}$.
From the data for the three system sizes, we evaluate its correlation range to be
$\Gamma_{ee}\sim 1$.
That is, a single renewal, after $\Delta\gamma 
\sim \Gamma_{pp}$, of the "active structures" is not sufficient for 
the non-affine field to fully decorrelate. This we take as a hint of 
the fact that active structures occupy on average a fraction only of 
the total system "volume". With this interpretation, we evaluate this 
fraction as $\Gamma_{pp}/\Gamma_{ee} \sim 1/4$.

\section{Zone emergence, flips and elastic couplings}
Two phenomenological models of plasticity of jammed disordered media -- 
namely the STZ theory of Falk and Langer~\cite{FalkLanger1998,FalkLanger2000}, and the soft glass 
rheology (SGR) of Sollich et {\it al}~\cite{SollichLequeuxH'ebraudCates1997,Sollich1998} -- have been 
proposed recently. Both describe, explicitly (STZ) or implicitly 
(SGR), plastic events as transitions concerning small regions, 
modelized as either Eshelby-like transformations or jumps 
out of traps in the energy landscape. These flips are viewed as independent, 
hence 
individual, and governed by the combined effect of external drive and 
of a thermal-like noise. 

A question then immediately arises : are our 
above results consistent with the basic ingredients of these models? 
We now try to answer it by careful inspection of the evolution of the 
spatial structure of the non-affine tangent field $\lbrace\bf{u}_{i}\rbrace$.

\subsection{Identification of active structures}
Let us first recall that $\lbrace\bf{u}_{i}\rbrace$ can be written as~\cite{MaloneyLemaitre2004a}:
\begin{equation}
\label{eq:hessien}
u_{i\alpha}= H_{i\alpha,j\beta}^{-1} \Xi_{j\beta}
\end{equation}
\begin{equation}
\label{eq:xi}
\Xi_{j\beta} = - \frac{\partial^{2} U}{\partial\gamma \partial 
r_{j\beta}}
\end{equation}
where $(i,j)$ label particles and $(\alpha,\beta)$ cartesian components.
$U$ is the total energy of the LJ system, and $\bar{\bar{H}}$ 
the associated hessian matrix. $\bf{\Xi} \delta\gamma$ is the field of 
particle force increments generated by an infinitesimal increment of 
homogeneous shear. $\lbrace\bf{u}_{i}\rbrace$ can be decomposed on 
the eigenmodes $\bf{\Psi}_{n}$ of $\bar{\bar{H}}$ (the phonon modes). Upon 
approaching a spinodal point, a single one, $\bf{\Psi}_{1}$, softens 
critically~\cite{MalandroLacks1999,MaloneyLemaitre2004a}. So, in near critical regions, 
$\lbrace\bf{u}_{i}\rbrace$ exhibits a square-root divergence and is 
dominated by its projection $\bf{u}_{1}$ on $\bf{\Psi}_{1}$.
This property enables us to clearly characterize the active structures close 
to the onset of plastic events.

The features we identify as characteristic of the evolution of 
$\lbrace\bf{u}\rbrace$ are exemplified on Figures 8 and 9. In these 
figures, for each value of $\gamma$ we decompose the total non-affine 
field as:
\begin{equation}
\label{eq:decompose}
\bf{u} = \bf{u}_{1} +  \bf{u}_{2} + \bf{\tilde{u}}
\end{equation}
with $\bf{u}_{1,2}$ its projections on the two modes $\bf{\Psi}_{1,2}$
with lowest eigenvalues ($\lambda_{1} < \lambda_{2}$). 

We observe that, in all cases, the active structures emerge out of a quasi-random small 
background as one or several localized zones characterized by strong 
quadrupolar contributions to $\bf{u}$. These 
quadrupoles are approximately aligned with the principal directions 
of the homogeneous strain. As zones approach instability, they soften, 
and the corresponding growing part of $\bf{u}$ concentrates into the 
lowest mode component $\bf{u}_{1}$.

\begin{figure}[p]
\begin{center}
\unitlength = 0.006\textwidth
\includegraphics[width=0.95\textwidth,clip]{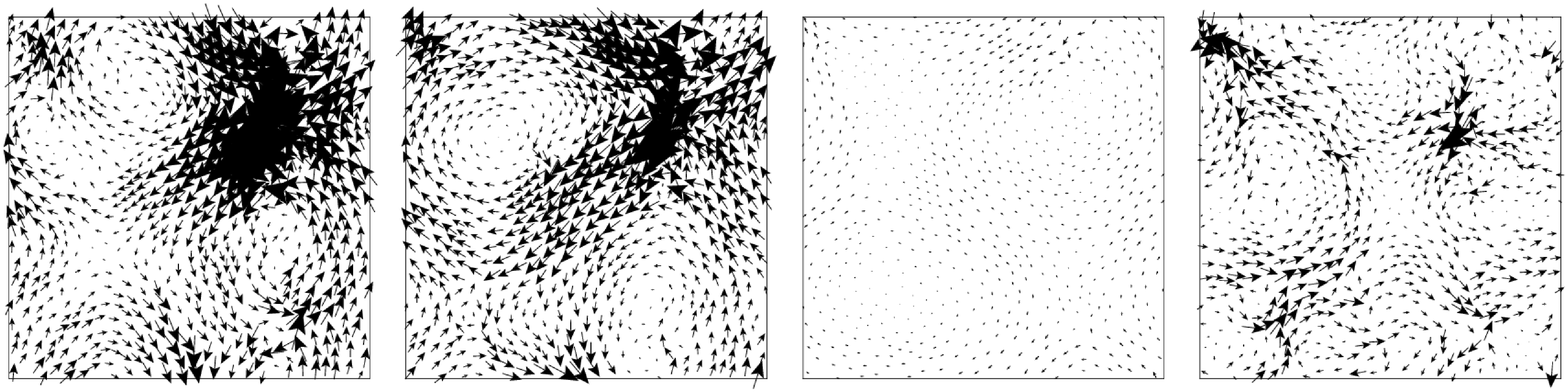}
\includegraphics[width=0.95\textwidth,clip]{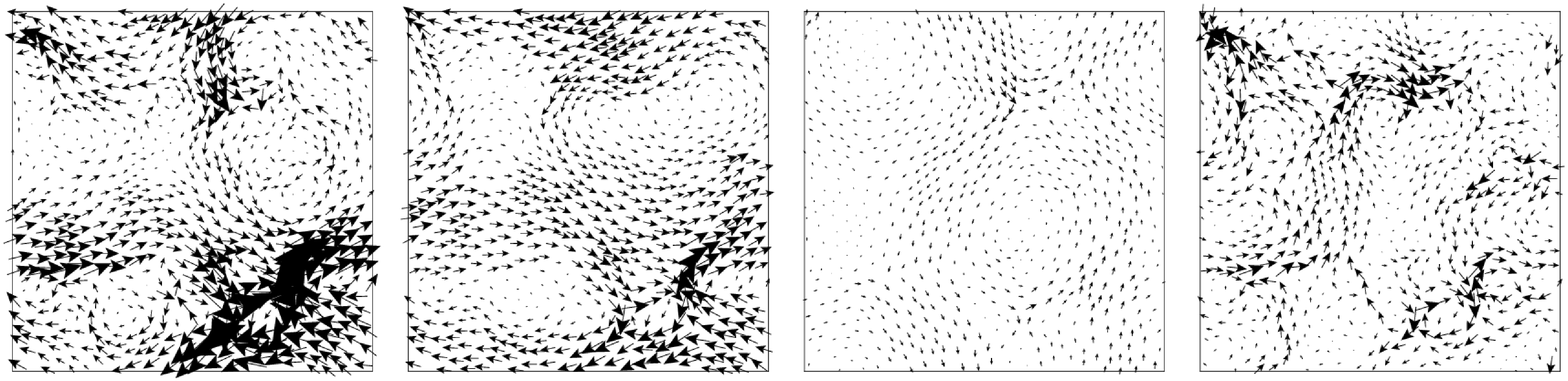}
\includegraphics[width=0.95\textwidth,clip]{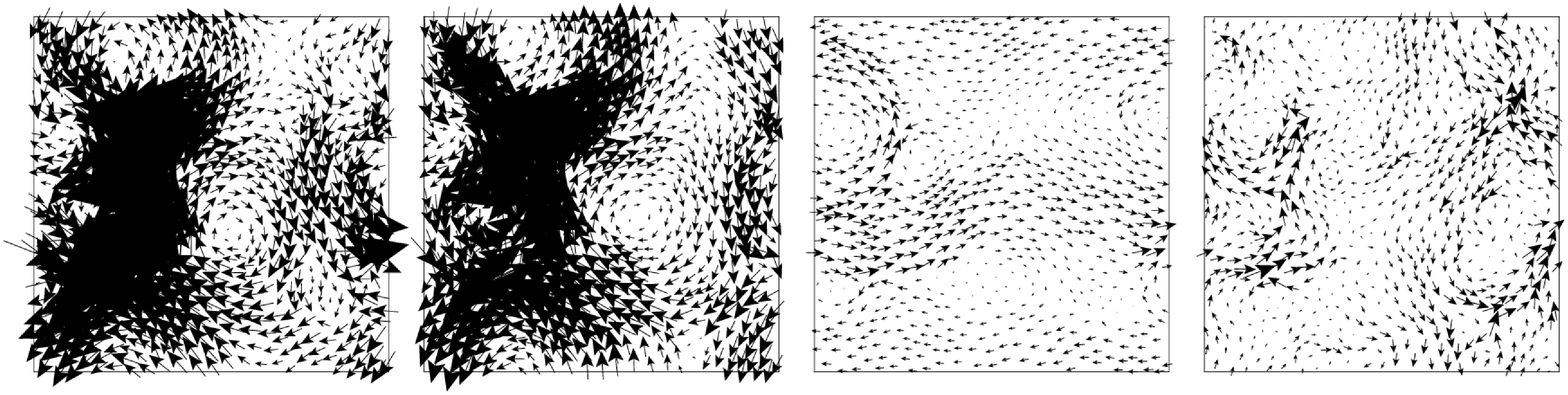}
\includegraphics[width=0.95\textwidth,clip]{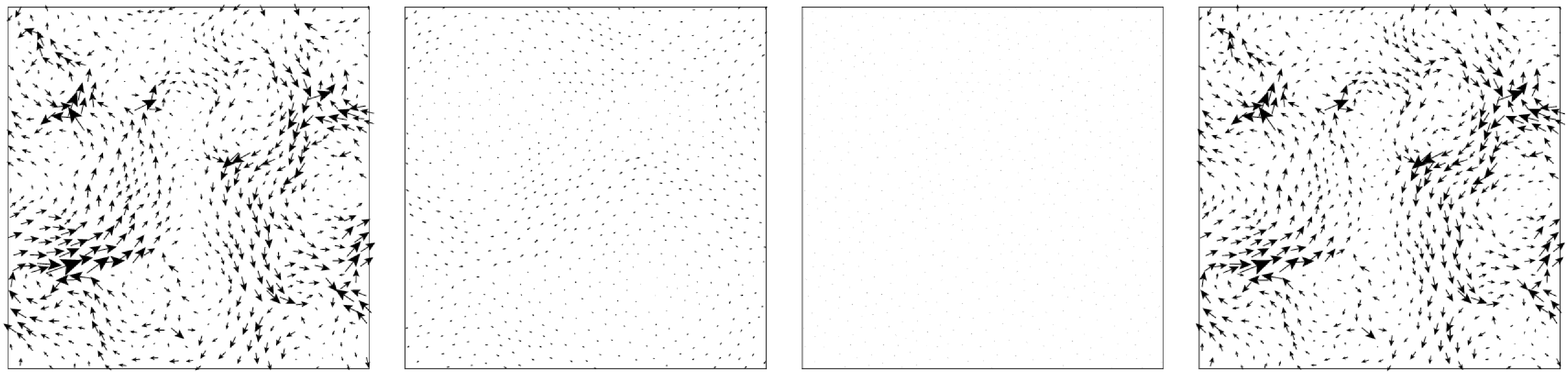}
\begin{picture}(100,0)(0,0)
\put(-31,32){\makebox(0,0){\large ${\bf d}$}}
\put(-31,74){\makebox(0,0){\large ${\bf c}$}}
\put(-31,117){\makebox(0,0){\large ${\bf b}$}}
\thicklines
{
\put(3,37){\circle{20}}
\put(118.5,37){\circle{20}}
\put(118.5,79){\circle{20}}
}
\thinlines
\put(-31,160){\makebox(0,0){\large ${\bf a}$}}
\end{picture}
\end{center}
\caption{\label{fig:556}
System size: $L=20$. Each line corresponds to a single strain value $\gamma$. 
The four frames present, from left to right, the non-affine tangent field $\bf{u}$ 
and its three components $\bf{u}_1$, $\bf{u}_2$, $\bf{\tilde u}$ as defined 
by equation~(\ref{eq:decompose}).\\
\mbox{}\hspace{2.5mm} (a): $\gamma=0.0556$: a quadrupole, clearly visible in the 
upper right corner of both $\bf{u}$ and the soft mode component $\bf{u}_1$, 
signals a near-critical zone Z, which flips at $\gamma_c\in[0.0557,0.0558]$.\\
\mbox{}\hspace{2.5mm} (b): $\gamma=0.0558$: Z has just flipped and disappeared, pushing another zone Z' 
(lowest right corners of $\bf{u}$ and $\bf{u}_1$) closer to its threshold.
Note that Z' was already discernable (line a) in both $\bf{u}$ and $\bf{\tilde u}$
before the event.\\
\mbox{}\hspace{2.5mm} (c-d): Zone Z, now indicated by circles, 
can be traced {\it back\/} to $\gamma=0.0269$ 
(line c, frame $\bf{\tilde u}$).
It survived the plastic event which occurred in the interval 0.0269 (line c)
and 0.0270 (line d).
}
\end{figure}

After identifying them in prespinodal regions, we are able to trace 
them {\it back} over sizeable $\gamma$-ranges before their amplitude has 
decreased enough for them to gradually merge into the global 
disordered background structures. In some cases, this 
range of visibility extends across one or more plastic events. 
A typical sequence illustrating this behavior, shown on Figure~\ref{fig:556},
extends over a range of $\Delta\gamma\sim3\%$, to be compared with the average
length $\Delta\gamma\sim 1\%$ of elastic episodes for this system size.

\begin{figure}[p]
\begin{center}
\unitlength = 0.006\textwidth
\includegraphics[width=0.95\textwidth,clip]{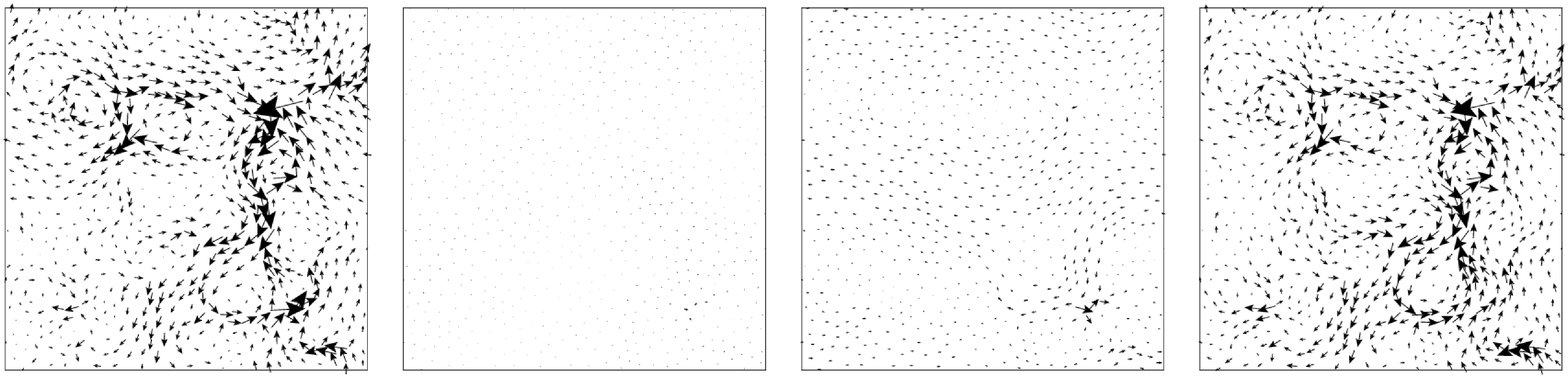}
\includegraphics[width=0.95\textwidth,clip]{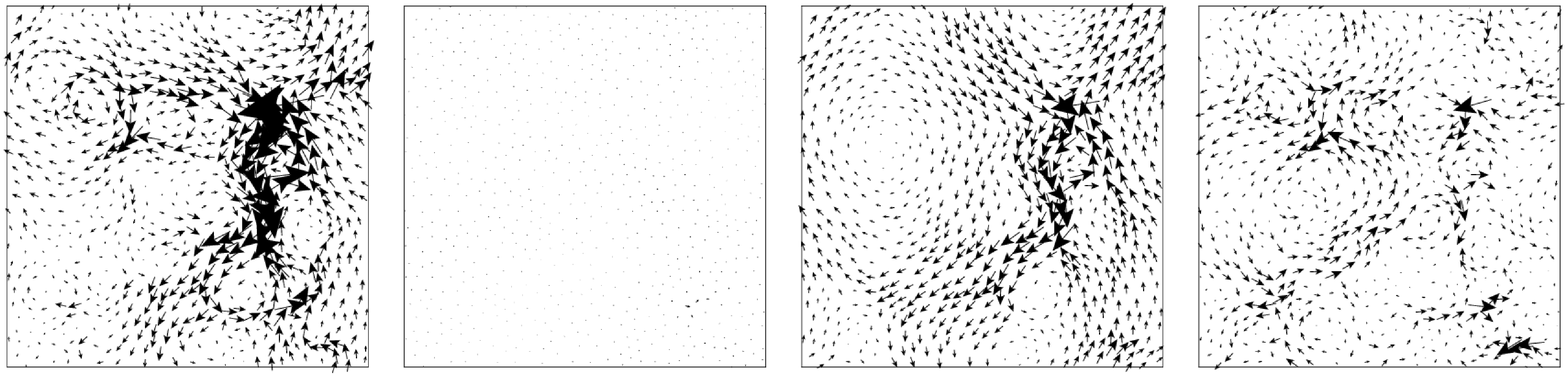}
\includegraphics[width=0.95\textwidth,clip]{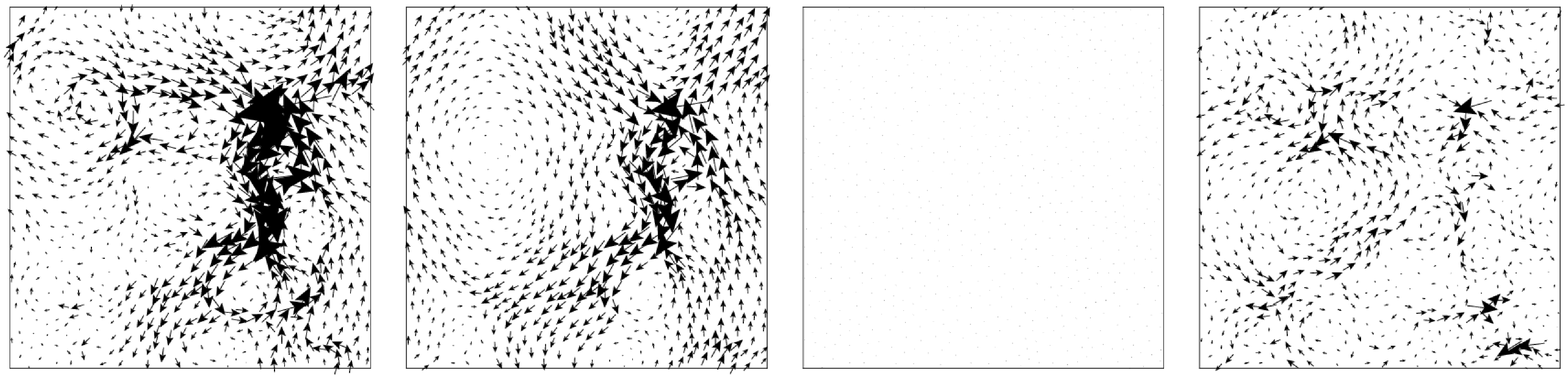}
\includegraphics[width=0.95\textwidth,clip]{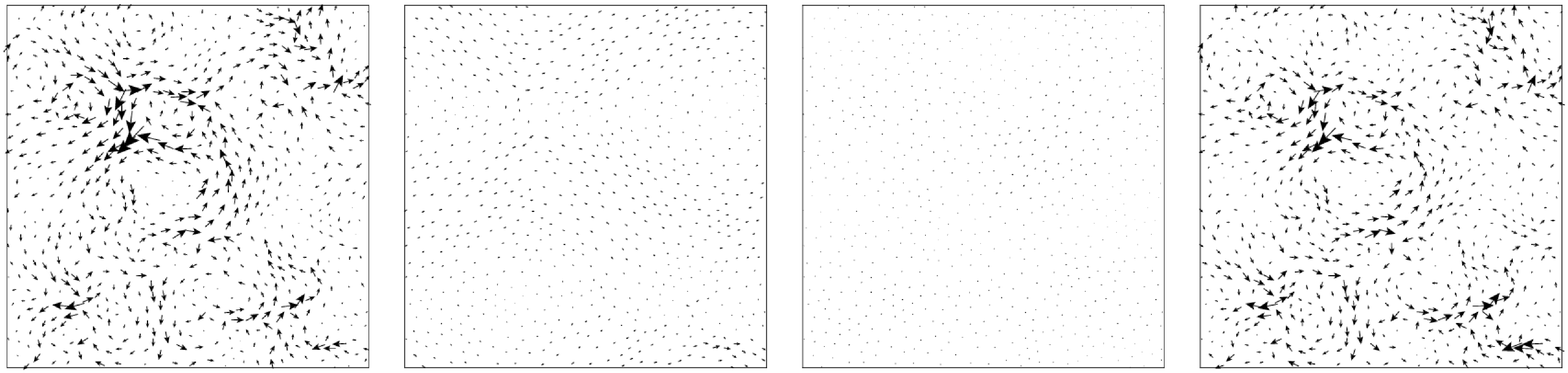}
\begin{picture}(100,0)(0,0)
\put(-31,32){\makebox(0,0){\large ${\bf d}$}}
\put(-31,74){\makebox(0,0){\large ${\bf c}$}}
\put(-31,117){\makebox(0,0){\large ${\bf b}$}}
\put(-31,160){\makebox(0,0){\large ${\bf a}$}}
\end{picture}
\end{center}
\caption{\label{fig:1150} 
From a to d: $\gamma =$ 0.1150, 0.1160, 0.1161, and 0.1166. 
Two vertically aligned zones flip together in the interval $0.01165-66$ and disappear.
As strain increases from $\gamma=0.1150$, they soften and sequentially appear in 
$\bf{u}_2$, then in $\bf{u}_1$ after a clear level crossing in the interval $0.01160-61$.
The connecting flow lines are a signature of their elastic coupling.
}
\end{figure}


\subsection{Zone flips}
Each discontinuous drop on the $\sigma(\gamma)$ curve is associated 
with the sudden disappearance of one or more of these soft zones 
(Figures~\ref{fig:556} and~\ref{fig:1150}). Noticeably, the resulting changes in 
$\lbrace\bf{u}_{i}\rbrace$ 
remain quite localized, leaving most of the other prominent structures 
of the non-affine pattern essentially unchanged. In particular,  
zones which were already clearly visible are commonly seen to survive 
the plastic event (see Figure~\ref{fig:556}).

We have shown in Section II.2 that particle jumps $\lbrace 
\Delta \bf{R}_{i}^{a}\rbrace$ in plastic events  are well correlated 
with the precritical $\lbrace\bf{u}_{i}\rbrace$ structure. This 
entails that the flip of a zone Z is associated primarily with a 
quadrupolar displacement field of finite amplitude centered on Z.
This supports the representation, proposed by Argon and coworkers~\cite{BulatovArgon1994a,ArgonBulatovMottSuter1995} and explicited by Picard et al~\cite{PicardAjdariLequeuxBocquet2004}, 
of the elementary dissipative process as the Eshelby-like 
shear transformation of a self generated inclusion involving a few 
particles, which is also the basis of the STZ theory.

\begin{figure}[h]
\begin{center}
\psfrag{p}{{$\Pi_{pl}$}}
\psfrag{y}{{$\Delta y$}}
\includegraphics[width=0.45\textwidth,clip]{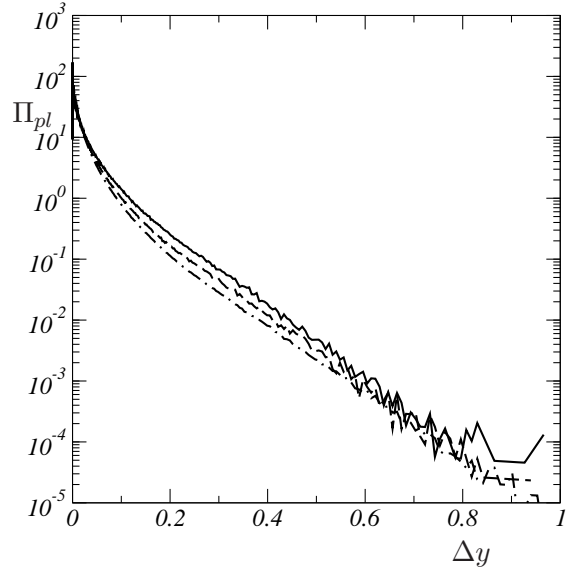}
\caption{\label{fig:tail}
The distribution $\Pi_{pl}(\Delta y, \delta)$ of particle displacements 
induced by plastic events for $L=10,20,40$.
}
\end{center}
\end{figure}

This representation is also consistent with the main features of the 
distribution $\Pi_{pl}(\Delta y, \delta)$ of particle displacements 
induced by plastic events (see $\S$ II.B). Indeed, the 
plot on Figure~\ref{fig:tail} shows that that $\Pi_{pl}$ becomes exponential for 
$\Delta y \gtrsim 0.2$. In the Eshelby picture, we expect large 
$\Delta y$ to correspond to displacements within the transforming 
zone(s), while smaller $\Delta y$ are associated with particles 
sitting in the surrounding elastic medium. In this picture, we 
interpret the exponential tail of $\Pi_{el}$, hence of the full 
distribution $P$, as reflecting the diversity of intrazone structures. 
We thus expect that, upon varying the system size, the logarithmic 
slope of the tail should remain constant. As shown on Figure~\ref{fig:tail}, this 
prediction is very nicely verified for our three $L$ values. Moreover, 
it appears that the tail amplitude decreases with increasing $L$. In 
our interpretation, the statistical weight of the exponential tail is 
controlled by the volume fraction of zone cores involved in plastic 
events. Since the average size of the avalanches constituting plastic 
events scales roughly as $L$, we expect the corresponding core volume fraction 
to decrease with size, in agreement with the observed behavior.

So, we interpret the existence of an exponential tail in $P(\Delta 
y)$ as a consequence of structural disorder within transforming 
zones, and by no means as a signature of avalanche behavior, which 
only affects the tail amplitude.

\subsection{Elastic couplings}
\begin{figure}[h]
\begin{center}
\unitlength = 0.006\textwidth
\includegraphics[width=0.95\textwidth,clip]{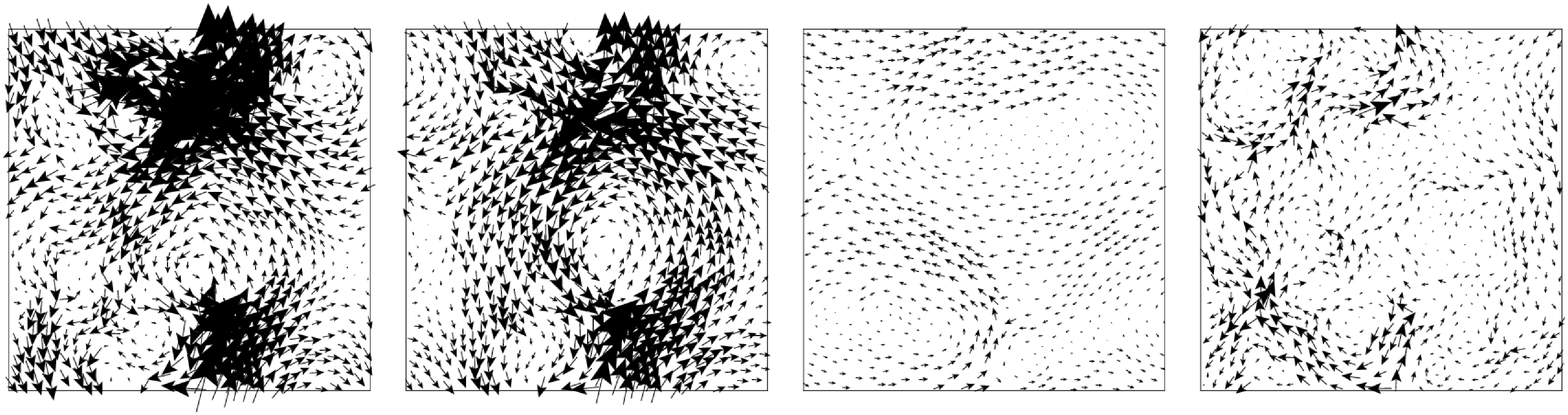}
\end{center}
\caption{\label{fig:265}
Two distant zones soften simultaneously and appear in $\bf{u}_1$.
The characteristic vortex-like structure signals their elastic coupling.
}
\end{figure}

Consider the sequence shown on Figure~\ref{fig:1150}. For $\gamma = 0.1150$ 
(Figure~\ref{fig:1150}-a) two 
zones are clearly discernible in the $\bf{u}$ pattern. 
At $\gamma = 0.1160$ (Figure~\ref{fig:1150}-b) their amplitude has grown, and they appear in the
projection $\bf{u}_{2}$ of $\bf{u}$ on the next-to-lowest mode.
At $\gamma = 0.1161$ (Figure~\ref{fig:1150}-c) they are slightly softer and  have invaded the 
projection $\bf{u}_{1}$ on the lowest mode. Note that, in these last 
two cases, they appear in the non-trivial soft mode as connected by 
"flow lines" which reproduce the most prominent vortex structures 
first described by Tanguy et al~\cite{TanguyWittmerLeonforteBarrat2002}. 
This is particularly clear when two relatively distant zones soften simultaneously 
as shown on Figure~\ref{fig:265}.

From these and many 
similar observations (see Figure~\ref{fig:556}) we deduce that :

$\bullet$ Soft zones are coupled elastically via their 
quadrupolar fields.

$\bullet$ It is the associated flow lines which form the conspicuous 
vortex-like patterns characteristic of non-affine fields in amorphous 
solids. That such patterns are akin to incompressible flows results 
from the fact that these systems are much more compliant in shear than 
in compression. For our 2D LJ glass~\cite{MaloneyLemaitre2006}: $\mu = 39$, 
$K = 236$.

Quadrupolar couplings result, in the homogeneous 
elastic continuum approximation for the background medium, from a stress field 
$\sim {\rm cos}(4\theta)/R^{2}$ (with $R$ the length and $\theta$ the 
orientation with respect to $\hat{x}$ of the interzone vector).
Notice that, in Figure~\ref{fig:1150}, in $\bf{u}_{1}$ the fields of the two zones 
$Z_{1}$, $Z_{2}$, which are quasi vertically aligned, combine 
positively. We have checked that, in consistence with the above 
remark, zone pairs lying in sectors corresponding to negative 
couplings combine negatively in the phonons $\bf{\Psi}_{1}$ and 
$\bf{\Psi}_{2}$: in this case, one of the two zones appears with an opposite
sign, i.e. with reversed flow lines. These observations confirm the relevance
of inter-zone elastic couplings.\\

It is intuitively clear that configurations with strongly coupled soft 
zones  are good candidates for simultaneous flips, as illustrated on 
Figure~\ref{fig:556} and~\ref{fig:1150}. We observe a number of multi-zone flips, involving the 
disappearance of at least one diverging quadrupole, as well 
as that of some less visible zones. Moreover, plastic events alter the 
amplitude of the near-critical surviving zones in $\bf{u}$ 
(see Figure~\ref{fig:556}) in a way 
which appears roughly consistent with quadrupolar couplings.

\section{A phenomenology of shear zones}

The various pieces of information which we have gathered and 
presented above can now be organized into a rather detailed 
phenomenology of plasticity in our system in the AQS regime, which
can be schematized as follows.

{\bf (i)} Structural disorder gives rise, in such a glassy 
system, to the existence of strong inhomogeneities, or zones. 
These zones can be viewed as small inclusions \`a la Eshelby, embedded 
in a quasi-homogeneous elastic background, and plastically transformable 
under shear. They are advected by external shear toward their 
instability threshold. Upon approaching it, a 
zone softens, then flips at its spinodal and disappears.

The correlation range $\Gamma_{pp}$ between plastic 
events (see Section II.2) provides an evaluation of the amount of strain  
necessary to fully renew the population of zones. Here 
$\Gamma_{pp} \sim 0.25$.

One renewal is not sufficient, however to decorrelate fully the non 
affine field. This suggests that zones occupy a fraction only of the 
system, of order $\Gamma_{pp}/\Gamma_{ee}$, here $\sim 1/4$.\\

{\bf (ii)} Zone elastic softening and plastic flipping are both associated with 
quadrupolar components in the non-affine displacement field which 
give rise to inter-zone elastic couplings. Due to the long range of 
elastic fields, a zone flip thus alters the local strain level at any 
other zone site in the system, the resulting $\gamma$-shift 
depending, in amplitude and sign, on the relative position of source 
and target. These signals have two types of effects.

--- A flip signal may shift the strain level of some zones beyond 
their threshold $\gamma_{c}$, thus triggering their flip and 
initiating a cascade. In the QS regime, where acoustic delays are 
neglected, such cascades are instantaneous.

--- For the non-flipping zones, the elastic signals resulting from 
flips constitute a dynamical noise, acting in parallel with the 
externally imposed advection, 
whose frequency scale is proportional to the strain rate $\dot{\gamma}$.

So, in agreement with the previous proposition of Bulatov and Argon~\cite{BulatovArgon1994a,ArgonBulatovMottSuter1995}, also underlying recent models by Picard et al~\cite{PicardAjdariLequeuxBocquet2005} and by Baret et al~\cite{BaretVandembroucqRoux2002}, we conclude that elastic 
couplings between zones play an essential role, by inducing avalanches 
and generating the dominant contribution to the disorder-induced noise.
At $T =0$ and moderate strain rates, it is this dynamical elastic noise 
which, in addition to strain advection, must appear in models of 
plasticity of jammed media. In a forthcoming article, we will propose 
such a schematic model, and show that it accounts for a system-size scaling
behavior of avalanches.\\

Within the foregoing picture, we interpret the two kind of events 
invoked by Tanguy et al~\cite{TanguyLeonforteBarrat2006} as plastic events of different 
sizes. Namely, their transient shear bands exhibit the characteristics 
expected for large avalanches of the type described above, their 
directionality resulting from that of quadrupolar couplings in their 
system with two rigid walls. Their local events correspond to the 
small stress drops appearing during initial transients (see Figure~\ref{fig:stressstrain}): 
at low stress levels, before advection has been able to significantly 
feed the near-threshold population, we indeed understand that 
avalanche triggering is unlikely, hence that single flip events are 
the rule.\\

Let us finally stress that the above zone phenomenology remains very  
schematic. Indeed, exhaustive inspection of the evolution of $\bf{u}$ 
reveals that zone life is somewhat more eventful
than our simplified description suggests. Rather frequently, 
we see a given zone undergoing a few flips before it disappears. Such 
zones can be termed multi-state. In other instances, a barely visible 
zone emerges fast enough to "overtake" previously more visible, 
softer, ones. That is, zone moduli and, very likely, thresholds are 
not unique, but distributed about an average. These remarks point 
toward the interest of pursuing extensive characterization of elastic 
heterogeneity in jammed systems, in particular via studies of coarse 
grained elastic moduli.\\ 

On a more speculative level, we would like to raise an important 
question: are the irreversible transformations identified here under 
shear related to the dynamic heterogeneities~\cite{DoliwaHeuer2003,DoliwaHeuer2003a,DoliwaHeuer2003b,SchroederSastryDyreGlotzer2000} characteristic of 
glassy dynamics near and below $T_{g}$?
Indeed, we share the opinion, formulated long ago by Goldstein~\cite{Goldstein1969}, 
that at finite temperature "rearrangements are of 
course occurring all the time in the absence of an external stress; 
the external stress, by biasing them, reveals their existence".

Our active zones are primarily sensitive to shear. This, we think, 
must be put together with recent results by Widmer-Cooper and
Harrowell~\cite{Widmer-CooperHarrowell2005,Widmer-CooperHarrowell2006}. 
These authors show that dynamic 
propensity in a 2D LJ glass is uncorrelated with free volume, which we 
understand to mean that their local rearranging structures are only 
wekly coupled to compression. This leads us to suggest that their 
observation that zones of 
high propensity have large "Debye-Waller factors" amounts to 
identifying them as soft zones sensitive to shear -- a speculation 
which will demand extensive future investigation.


\begin{thebibliography}{10}
\bibitem{footnote}
Note that the position of the peak of $\tilde{\Pi}_{el}$ 
corresponds, roughly, to the width of the gaussian approximant of the 
small-$\Delta y$ part of $P(y, \delta)$.
\bibitem{footnote2}
Strictly speaking, in order to observe the asymptotic $\Delta\gamma^{1/2}$ behavior,
we would need data for decreasing simulation strain intervals $\delta\to0$. 
With our value $\delta=10^{-4}$, we measure at small $\Delta\gamma$ an effective
behavior $D_{ep}\sim \Delta\gamma^{0.6}$.
\end{thebibliography}

\end{document}